# Strategic Roadmap for Quantum- Resistant Security: A Framework for Preparing Industries for the Quantum Threat

Arit Kumar Bishwas[1,*]
PricewaterhouseCoopers
Fremont, USA
arit.kumar.bishwas@pwc.com

Mousumi Sen[2]
Independent Researcher
Fremont, USA
mousumisen.official@gmail.com

**Abstract—As quantum computing continues to advance, its ability to compromise widely used cryptographic systems projects a significant challenge to modern cybersecurity. This paper outlines a strategic roadmap for industries to anticipate and mitigate the risks posed by quantum attacks. Our study explores the development of a quantum-resistant cryptographic solutioning framework for the industry, offering a practical and strategic approach to mitigating quantum attacks. We, here, propose a novel strategic framework, coined name "STL-QCRYPTO", outlines tailored, industry-specific methodologies to implement quantum-safe security systems, ensuring long-term protection against the disruptive potential of quantum computing. The following fourteen high-risk sectors—Financial Services, Banking, Healthcare, Critical Infrastructure, Government & Defence, E-commerce, Energy & Utilities, Automotive & Transportation, Cloud Computing & Data Storage, Insurance, Internet & Telecommunications, Blockchain Applications, Metaverse Applications, and Multiagent AI Systems—are critically assessed for their vulnerability to quantum threats. The evaluation emphasizes practical approaches for the deployment of quantum-safe security systems to safeguard these industries against emerging quantum-enabled cyber risks. Additionally, the paper addresses the technical, operational, and regulatory hurdles associated with adopting quantum-resistant technologies. By presenting a structured timeline and actionable recommendations, this roadmap with proposed framework prepares industries with the essential strategy to safeguard their potential security threats in the quantum computing era.**

## I. Introduction

We are living in an era of remarkable innovation, where digital advancements are transforming our daily lives and reshaping the way we experience the world. From the cloud **[1]** to artificial intelligence **[2]**, and now quantum computing **[3]**, technological progress is accelerating at an unprecedented pace. With each breakthrough, new opportunities emerge, offering the potential to revolutionize industries, enhance the human experience, and solve complex global challenges.

Quantum computing, in particular, is opening doors to the future, enabling the design of advanced AI systems in near future, the discovery of new materials **[4]**, and the acceleration of drug discovery **[5]**. It is also driving new possibilities in environmental exploration, leveraging more sophisticated quantum-based sensors **[6]** for deeper insights. These developments represent not just the next phase of technological growth but a leap towards a more intelligent and connected future.

As exciting as these advancements are, they also introduce new challenges. As we embrace these transformative technologies, we must be prepared to navigate the complexities they bring, ensuring that we harness their full potential while addressing security and ethical considerations. The emergence of fully functional quantum systems, particularly those with a few hundred error-free qubits, poses a significant threat to the integrity of modern digital cybersecurity. One of the greatest concerns is the potential collapse of current encryption methods **[7]**, which rely on complex mathematical algorithms, especially those based on prime factorization **[8]**.

This threat was first demonstrated in 1996 when Peter Shor introduced his quantum algorithm **[9]**, showing that quantum computers could break cryptographic systems designed on classical mathematics, such as RSA encryption **[10]**. Though discussions around quantum attacks and countermeasures have been ongoing for decades, it was Shor's algorithm that underscored the urgency of addressing these vulnerabilities. His work marked a turning point, driving global awareness of the need to develop quantum-resistant cryptographic solutions to safeguard the future of digital security. As quantum technology evolves, businesses must prioritize quantum-safe encryption methods to protect against this impending cybersecurity challenge.

While the progress in quantum computing is indeed impressive **[11][12]**, it may still be some time before fully functional quantum computers are commercially available at scale. However, this doesn't mean we can afford to ignore the potential threats they pose. In fact, it's crucial to adopt an "act-now" approach to cybersecurity. Consider this scenario: if your sensitive data were stolen today, it might remain encrypted and seemingly safe. But years from now, when mature quantum systems become available, that same data could be decrypted, leaving you with no defence. Quantum computers have the potential to break many of today's cryptographic algorithms, putting long-term sensitive data at significant risk. Therefore, businesses must start preparing now by adopting quantum-resistant encryption methods to protect against future vulnerabilities. Proactive measures today will be essential to safeguard critical information in the quantum era.

In this paper, we present a complete strategic framework for mitigating quantum attacks in industrial settings. At first part of the framework, our approach is structured into three distinct strategic levels: foundational level, intermediate level, and advanced level. These levels are designed based on the priority of implementation and the complexity of execution, providing a comprehensive pathway for organizations to strengthen their defence mechanisms against quantum threats. The next part of the framework includes a process of adoption and implementation applied to each strategic transition level. Additionally, we explore multiple potential industrial use cases aligned with each approach level, offering practical insights into their application. Furthermore, we emphasize the regulatory considerations that industries must address when implementing mitigation strategies for quantum attacks, ensuring compliance with relevant legal and policy frameworks.

## II. Security Landscape and Quantum Computing

In recent years, the field of quantum computing has made remarkable progress, showcasing its potential for exponential speedup compared to classical computing systems **[13][14][15][16].** This groundbreaking technology has the potential to revolutionize the way we approach a wide range of fields, from solving complex optimization and artificial intelligence challenges to advancing material design and accelerating drug discovery. However, there is a darker side too which is scary to think as these advances also pose a significant threat to current cryptographic systems, as classical security relies on communication channels that could be easily compromised by a sufficiently powerful quantum computer. Quantum cryptography offers a solution to this vulnerability, as it provides a quantum-secure method of key distribution, making it nearly impossible to intercept communication channels **[17]**.

The foundation of quantum cryptography dates back to 1970, when Stephen Wiesner **[18]** introduced the concept of quantum money, which inspired further research into quantum key distribution (QKD). Leveraging principles from the Einstein-Podolsky-Rosen (EPR) experiment **[19]**, the field gained momentum in the 1980s and 1990s with Bennett and Brassard's landmark BB84 protocol **[20]** in 1984. This breakthrough demonstrated the feasibility of secure key distribution via quantum channels using photon polarization. In 1991, Artur K. Ekert **[21]** advanced this work with a variation of BB84, incorporating Bell states and quantum entanglement, drawing on Bell's theorem and the Clauser-Horne-Shimony-Holt inequalities **[22]**. In 1992, the BBM92 protocol **[23]**, an extension of BB84, was introduced and remains a foundational resource for the proposed quantum cryptographic framework.

In our daily lives, we are familiar with classical mechanics, which is relatively easy to comprehend. However, quantum mechanics operates at the atomic level and exhibits some unusual phenomena. Quantum computers are built on these unique quantum mechanical properties. While classical computers use bits, where each bit is either in the state of "0" or "1," quantum computers use qubits, which can exist in both states, "0" and "1," simultaneously. A quantum computer basically works based on the following key essential quantum mechanical principles:

Superposition: In quantum mechanics, superposition refers to a quantum state being in more than one state at the same time. Each state in superposition has a certain probability of existing, allowing quantum computers to process multiple possibilities simultaneously **[24]**.

Entanglement: In quantum physics, entanglement occurs when two or more particles become linked in such a way that a change in one particle instantly affects the others, no matter how far apart they are. This property was used by Ekert in 1991 to develop the quantum key distribution (QKD) algorithm, Ekert91. Another entanglement-based QKD algorithm, BBM92, was proposed in 1992. The exact cause of entanglement remains an open question in physics **[25]**.

Interference: It is a fundamental concept in quantum mechanics, where it occurs when quantum waves, or wavefunctions, overlap and combine. These waves represent the probabilities of different quantum states, and when they interact, they can either reinforce or cancel each other out. This phenomenon is known as constructive interference when the waves amplify each other, or destructive interference when they cancel each other out. In quantum computing, interference plays a key role in controlling qubits and optimizing quantum algorithms. By harnessing the principles of interference, quantum computers can manipulate probabilities in a way that classical computers cannot, allowing for more efficient problem-solving **[26]**.

No-cloning theorem: This principle states that quantum states cannot be copied. Once we interact with a quantum system, its superposition collapses into a single, classical state **[27]**.

Decoherence: Although quantum states can exist in superposition, when we measure or interact with the system, the superposition collapses, and the system settles into a single state. As a result, we cannot measure multiple quantum states at once **[28]**.

## III. Strategic Planning & Adoption: Steps Toward Quantum-Safe Security

The quantum threat to current security systems is significant and requires an immediate "act-now" approach across industries. Given the complexity of existing classical security infrastructure setup, a strategic and well-planned approach is essential to address potential quantum attacks effectively. Recognizing the intricate dependencies within today's security infrastructure, we have developed a quantum-safe security framework to proactively address and mitigate future risks, ensuring organizations are prepared for the evolving threat landscape. Cryptography is generally divided into two main categories: symmetric and asymmetric cryptography:

### Asymmetric cryptography

Asymmetric cryptography **[29]** relies on complex mathematical calculations, with RSA being one of the most widely used algorithms. RSA's security is based on the difficulty of factoring large numbers into their prime components, a problem considered computationally intractable. While asymmetric cryptography offers strong security, it is significantly slower than symmetric methods and requires much larger keys to maintain similar levels of protection. If faster factorization techniques are developed, it would necessitate even larger prime numbers, increasing computational load and slowing down processes. Asymmetric systems use two keys: a public key for encryption and a private key for decryption, making them less efficient for widespread use in today's fast-paced digital environment, **Fig 1** demonstrates the encryption and decryption in asymmetric cryptography.

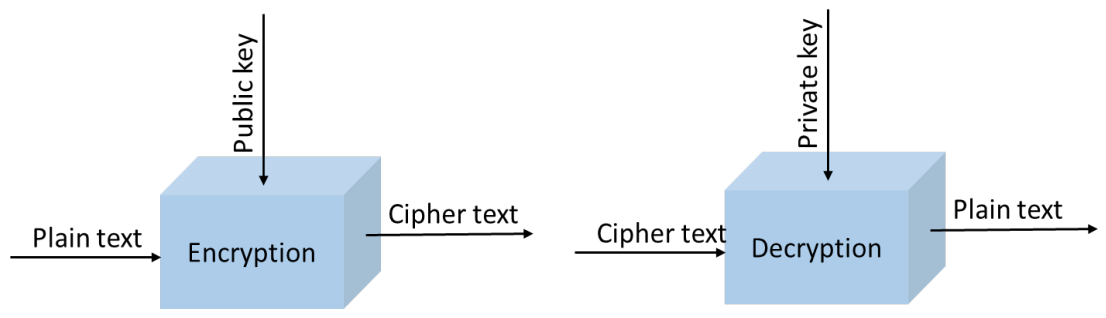

Fig 1: Encryption and decryption

### Symmetric cryptography

In symmetric cryptography **[30]**, the same secret key is shared between the sender and receiver via a secure, authenticated channel. The key is typically generated using a pseudorandom number generator, though true randomness is challenging to achieve. Careful selection of a high-entropy random number generator is crucial. Symmetric cryptography is faster and requires shorter key lengths compared to asymmetric methods while providing strong security. One of the most widely used and secure symmetric algorithms is AES (Advanced Encryption Standard) **[31]**, which supports key lengths of 128, 192, and 256 bits for both encryption and decryption, shown in **Fig 2**.

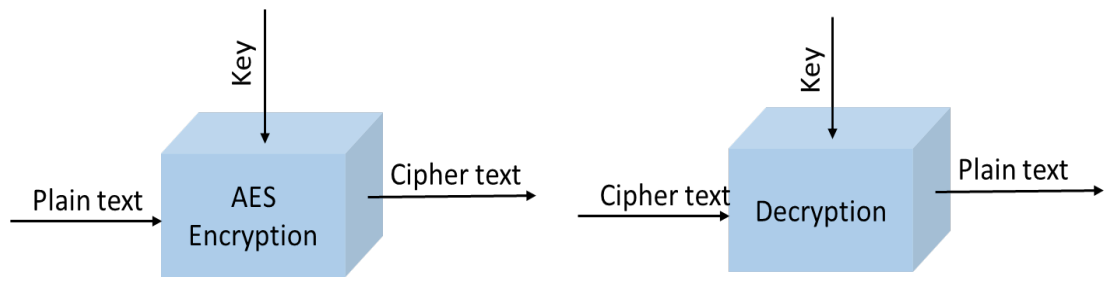

Fig 2: Encryption and decryption with AES protocol

From a strategic perspective, we have defined three strategic levels to guide the transition from a classical-only security approach to a quantum-safe solutioning strategy within the security landscape. These levels provide a clear, phased roadmap for organizations to gradually adopt quantum-resistant solutions while minimizing operational disruptions. By addressing immediate vulnerabilities, planning for mid-term upgrades, and preparing for long-term quantum innovations, this framework ensures a proactive and comprehensive response to emerging quantum threats. Implementing this multi-level strategy will enable businesses to future-proof their security infrastructure against the evolving technological landscape.

In our strategic roadmap for adopting quantum-safe security, we introduce **"strategic transition level" (STL)** framework which outlines three key levels of defence. **STL-1** (foundational level) focuses on securing systems using asymmetric cryptography in anticipation of quantum threats. **STL-2** (intermediate level) and **STL-3** (advanced level) address more advanced security measures, emphasizing symmetric cryptography strategies to safeguard against potential quantum attacks. This multi-tiered approach ensures comprehensive protection as the quantum landscape evolves.

### A. *Strategic Transition Level 1 (STL-1): Foundation - Classical way Post-Quantum Safety*

Recently, NIST released first post-quantum cryptography standards **[32]** which enables the industry to think to "act-now". Over the time, researchers keep on exploring and designing new classical cryptographic algorithms that can withstand quantum attacks, noticeably are latticed based and hash-based algorithms. According to NIST recent announcement, it has standardized three cryptographic algorithms, with which post-quantum cryptography will be secure against attacks from classical and quantum computers. Currently, there isn't a large enough quantum computer that threatens the current level of security, but Moody **[33]** said that agencies need to be prepared ahead of future attacks.

NIST has chosen CRYSTAL-Kyber **[34]** as the primary key encapsulation mechanism (KEM) for securing encryption processes, such as those used for public-facing websites. Additionally, three digital signature algorithms—CRYSTAL-Dilithium **[35]**, FALCON **[36]**, and SPHINCS+ **[37]**—have been selected for standardization (**Fig 3**). CRYSTAL-Dilithium, FALCON, and SPHINCS+ will be employed in situations where digital signatures are required to ensure authentication and integrity. These selections represent the industry's move toward quantum-resistant cryptographic standards to protect sensitive data in the evolving digital landscape.

We strongly recommend adopting the STL-1 approach as the immediate priority to secure sensitive data. This approach is highly cost-effective and straightforward to implement, offering a quicker and smoother transition compared to the more complex STL-2 and STL-3 strategies. It requires minimal changes to the existing infrastructure and has a negligible impact on the current cryptographic systems, making it an ideal starting point for organizations looking to enhance their security without significant disruption. By addressing immediate vulnerabilities while maintaining operational efficiency, the STL-1 approach provides a solid foundation for future security enhancements.

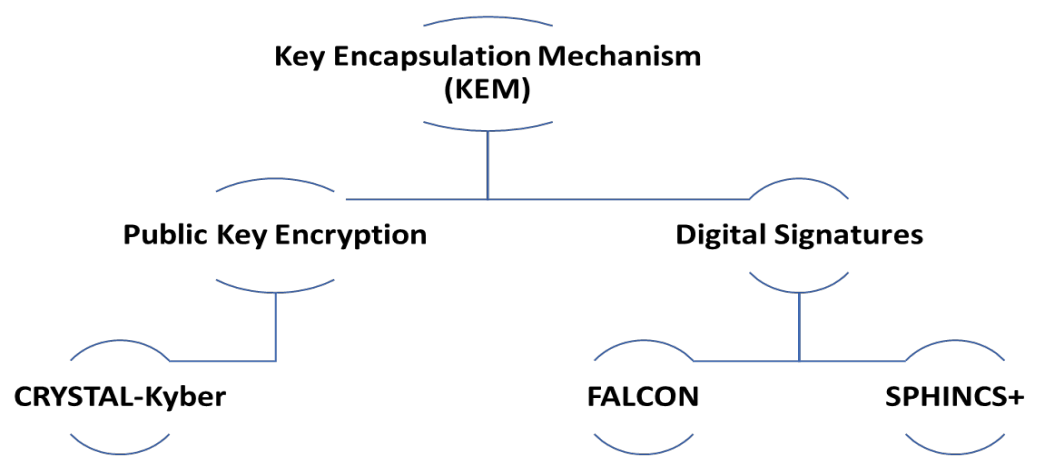

Fig 3: Primary key encapsulation mechanism recommended by NIST

### B. *Strategic Transition Level 2 (STL-2): Intermediate - Quantum Enhanced Approach*

At present, we do not have sufficiently mature quantum systems to fully test the STL-1 security strategy. In STL-1, we discussed using advanced classical algorithms based on solid mathematical theories to mitigate future quantum attacks. However, this approach may not offer complete protection, as more advanced quantum techniques could potentially breach the STL-1 defences in the future. To ensure long-term security, we recommend transitioning to the STL-2 strategy after implementing STL-1.

The STL-2 strategy emphasizes a hybrid quantum-classical approach, where Quantum Random Number Generators (QRNGs) **[38]** play a critical role. QRNGs leverage quantum mechanics to produce truly random numbers, unlike classical random number generators that depend on deterministic algorithms or physical processes. By incorporating QRNG-generated keys with classical cryptographic algorithms (such as "Advanced Encryption Standard" as an example **[39]**), the STL-2 strategy offers a more robust defence against post-quantum attacks. This hybrid approach significantly enhances security, providing better resistance to potential future quantum threats compared to the STL-1 strategy. Additionally, it allows businesses to gradually integrate quantum technologies while maintaining the effectiveness of their current systems.

While it's important to account for the costs associated with implementing QRNG devices—

including hardware expenses, training, and ongoing maintenance, etc.—these investments are higher compared to the STL-1 strategy. However, the increased cost is justified by the significantly enhanced security that QRNGs provide. Given the potential risk of future quantum-based security breaches, the investment in QRNG technology is a proactive measure that can safeguard critical assets and reduce the likelihood of costly security vulnerabilities. In the long run, the value of preventing quantum attacks far outweighs the upfront and operational expenses, making it a strategic and necessary investment for businesses focused on long-term cybersecurity resilience.

### C. *Strategic Transition Level 3 (STL-3): Advanced - Quantum Only Approach*

Although, in a quantum-classical hybrid approach, QRNG provides enhanced level of security for quantum attacks but it is still not safe enough as the classical part in this hybrid approach is still vulnerable. So, we need a pure quantum solutioning approach, so recommending the next level of strategy to adopt is STL-3 where we emphasis on Quantum Key Distribution (QKD) **[40]**, which is widely regarded as the most practical solution to mitigate quantum attacks, relying on the principles of quantum physics rather than complex mathematical algorithms. Often synonymous with quantum cryptography, QKD can be viewed as an advanced extension of symmetric cryptography, incorporating the foundational elements of quantum mechanics, **Fig 4** shows the QKD mechanism. By leveraging quantum properties such as superposition and entanglement, QKD ensures secure key exchange, making it highly effective in safeguarding sensitive data during transmission. This approach offers a robust defence against both current and future quantum threats, providing unparalleled security assurance.

The STL-3 strategy involves a complex adoption process and significant infrastructure transition costs. A thorough feasibility analysis is essential to identify system gaps and prioritize the transition from classical to quantum solutions effectively. Rigorous planning is crucial to ensure a smooth and cost-efficient transition and implementation.

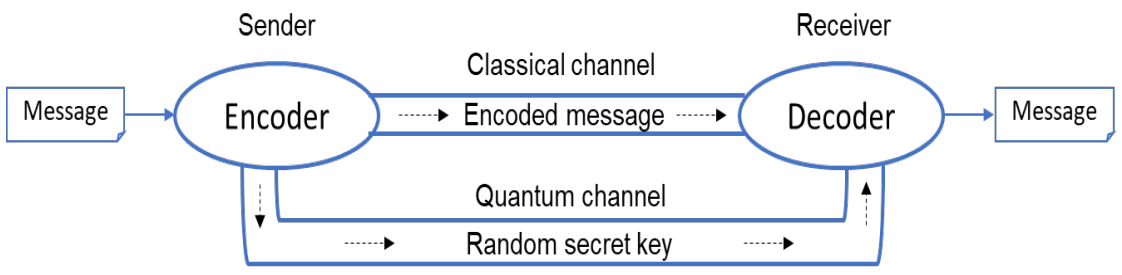

Fig 4: Quantum key distribution (QKD)

The following **Fig 5** highlights the strategic framework and associated levels for transition from classical to quantum security eco-system.

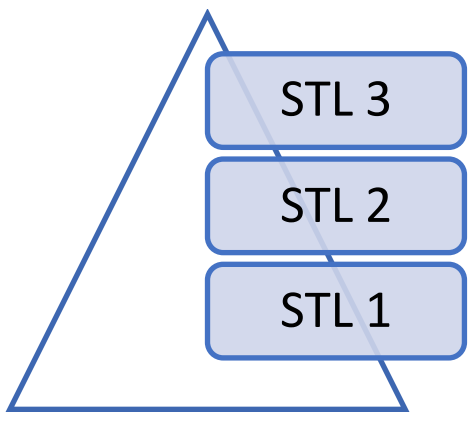

Fig 5: Strategic transition level framework

The following **Table 1** shares some of our proposed important parameters which helps in determining the feasible strategic level we should focus on to consider.

Table 1: Parameters to determine the strategic level we should focus on

| STL | Priorities | Transition Efforts | Implementation Complexity | Transition Cost | Special Skill Requirement | Adoption Timeline | Quantum safe |
|---|---|---|---|---|---|---|---|
| 1 | High | Low | Low | Low | Low | Immediate | Moderate |
| 2 | Medium | Medium | Medium | Medium | Medium | 0 - 1 year | Intermediate |
| 3 | Medium | High | High | High | High | 1 - 2 years | Advance |

## IV. Strategic Roadmap for Quantum-Resilient Industrial Applications

Quantum attacks present significant challenges across industries, but the greater hurdle lies in deploying a well-structured framework that strategically integrates quantum-resistant solutions into existing classical cryptographic infrastructures. To address this, we introduce our second novel and comprehensive strategic framework, **"QCRYPTO"**, designed to enable seamless quantum infusion into (or transformation from) current cryptographic systems.

The proposed QCRYPTO framework provides a methodical, industry-focused roadmap to help organizations transition from classical cryptographic methods to quantum-resilient solutions. Its high-impact methodology ensures that industries across sectors, including finance, healthcare, and critical infrastructure, are equipped to anticipate and mitigate quantum threats. By aligning with industry-specific requirements, our strategic QCRYPTO framework facilitates a smooth, phased adoption of quantum technologies, optimizing operational security while maintaining compliance with emerging global standards. This approach is critical as industries face the pressing need for quantum preparedness, ensuring that cybersecurity measures remain robust in the face of future quantum advancements.

### *A. Framework for Quantum-Resilient Technology Adoption*

The proposed strategic QCRYPTO framework is a seven-stage, strategic roadmap designed to guide industries in the adoption of quantum-resistant cryptography. Each stage builds upon the last, ensuring a methodical and high-impact transition toward quantum-resilient security. The framework is structured as follows:

**Quest**: This stage initiates the exploration of quantum threats and opportunities, driving industries to investigate the current landscape and emerging quantum technologies.

**Commence**: Here, the organization identifies vulnerabilities within its classical cryptography systems that are susceptible to quantum attacks, laying the groundwork for targeted intervention.

**Review**: At this stage, businesses evaluate their current infrastructure and risk levels, prioritizing areas for immediate improvement based on operational needs and the severity of potential quantum threats.

**Yield**: This involves the gradual integration of quantum-resistant solutions, ensuring compatibility with existing systems and fostering a culture of innovation and adaptability within the organization.

**Pivot**: A strategic overhaul of the organization's cryptographic architecture takes place, incorporating advanced quantum-resistant algorithms and enhancing security protocols in critical areas.

**Transcend**: In the final stage, industries scale their quantum-resilient security measures, amplifying defences across all digital assets, ensuring comprehensive protection, and preparing for future quantum advancements.

**Observe:** After implementing the solution, it is essential to continuously monitor and assess it for any security vulnerabilities. Regular updates and patches should be applied as advancements in technology emerge and new threats are identified. This proactive approach ensures that the solution remains resilient and effective against evolving cybersecurity challenges.

This methodical progression enables industries to not only secure their current systems but to future-proof their operations in a rapidly evolving quantum landscape, ensuring long-term resilience and competitive advantage. The **Fig 6** shows the strategic approach of the QCYPTO framework. By integrating the two frameworks **QCYPTO** and **STL**, we arrive at a comprehensive strategic solutioning framework named **"STL-QCRYPTO"**, shown in **Fig 7**.

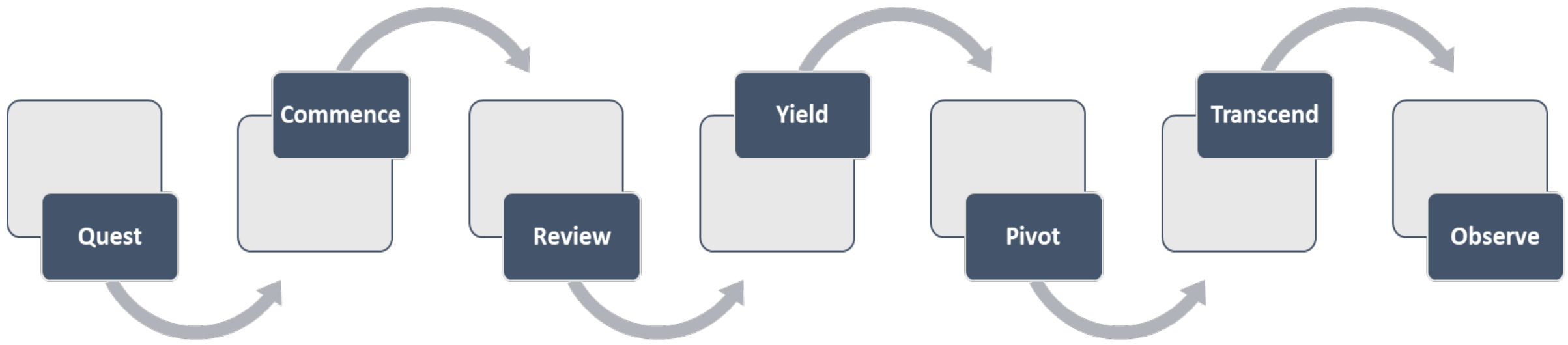

Fig 6: Strategic QCRYPTO framework

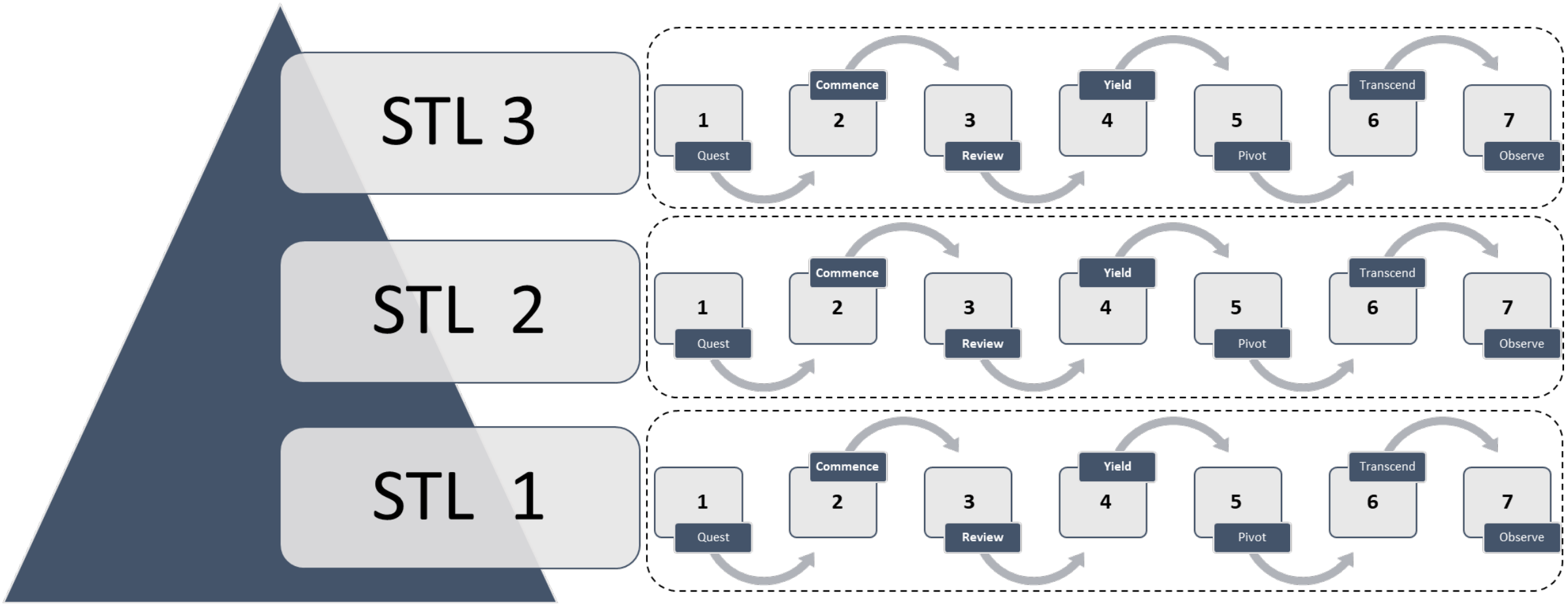

Fig 7: "STL-QCRYPTO" framework, a comprehensive complete strategic & solutioning framework to handle the quantum attacks.

*B. Infrastructure Roadmap*

Adapting current cryptographic infrastructures to withstand attacks from quantum computers, we need to consider and incorporate the hardware management and strategy for implementing which vary significantly depending on the size and complexity of the organization. Below is a breakdown of the hardware management requirements and strategic approaches for small, mid-size, and large organizations. From the **basic key management systems prospects, Hardware Security Module (HSM) [41]** is a dedicated hardware mechanism designed to securely manage, store, and use cryptographic keys. HSMs provide a high level of physical and logical security, ensuring that sensitive information, like encryption keys and certificates, cannot be accessed by unauthorized users or systems. Following are the key functions of HSMs:

- **Key Management**: HSMs generate, store, and manage encryption keys in a highly secure environment.
- **Data Encryption/Decryption**: HSMs handle cryptographic operations, such as encryption and decryption, without exposing the keys to other systems.
- **Digital Signatures and Authentication**: HSMs are used to sign and verify digital transactions securely.
- **Compliance and Certification**: Many industries use HSMs to meet regulatory compliance, like PCI-DSS in financial services, where cryptographic key management is essential.

In industrial terms, organizations can be categorized based on size and complexity of their IT infrastructure and security needs. This classification reflects the varying scales of technological and security needs across different business sizes, influencing strategic decisions in IT management and cybersecurity implementation:

- **Small Organizations** (fewer than 100 employees) typically operate with limited IT infrastructure and have modest security requirements. Their operations are often simpler, and they may rely on cloud-based or outsourced IT solutions.
- **Mid-Size Organizations** (100-1,000 employees) have more developed IT infrastructures and face increasingly complex security challenges. They often handle sensitive data, such as financial or healthcare information, and need more robust systems to ensure data protection and regulatory compliance.
- **Large Organizations** (more than 1,000 employees) manage highly complex, global IT infrastructures that include significant data centers and advanced security requirements driven by compliance mandates. These organizations need extensive resources to safeguard critical data and operations across multiple regions.

Some additional important considerations across all organization sizes:

- **Vendor Collaboration**: Collaborate with hardware vendors that are already developing quantum-ready hardware solutions.
- **Compliance & Certification**: Ensure that hardware solutions for PQC meet regulatory standards and certifications, such as NIST's post-quantum cryptography guidelines.

- **Scalability**: Hardware strategies must account for scalability as PQC algorithms become more widely adopted.

*1) Infrastructure for STL-1 Strategic Approach*

The current infrastructure for implementing post-quantum cryptography algorithms requires significant advancements, particularly in hardware scalability, seamless integration with existing cryptographic systems, and ensuring robust security for critical applications. Progress in developing post-quantum algorithms must move swiftly beyond mainstream cryptographic approaches to offer a quantum-resistant advantage for modern digital infrastructure. This will provide the necessary security foundation as organizations prepare for the quantum computing era, ensuring long-term protection against future threats.

*a) Infrastructure Compatibility with Hybrid Systems*

Hybrid cryptographic systems that combine classical cryptography with post-quantum algorithms allow a gradual transition, ensuring that systems remain secure during the shift to quantum-resistant security. Hardware such as security modules and cryptographic accelerators must be updated to support both classical and post-quantum cryptographic operations **[41][42]**. Research emphasizes the need for HSMs to be upgraded or replaced to support PQC. These HSMs will need to be capable of processing the larger key sizes and more complex computations required by PQC algorithms like CRYSTALS-Kyber or FALCON **[41]**.

*b) Computational Consideration*

Post-quantum algorithms often have higher computational requirements compared to classical cryptographic systems. CRYSTALS-Dilithium, for example, demands more computing power for key generation and encryption. Infrastructure must scale appropriately, with organizations potentially needing to upgrade processing power, memory, and storage to handle the increased load. Research suggests optimizing existing data centers and cloud infrastructures to handle PQC without significantly increasing latency. This is particularly important for sectors like financial services, where real-time processing and encryption are critical **[42]**.

*c) Standardization*

Existing network protocols and infrastructures will need to be updated to accommodate PQC algorithms. For example, implementing PQC in TLS (Transport Layer Security) requires careful handling of key exchange mechanisms, which differ significantly from current standards. Recent studies indicate that network routers, switches, and communication protocols must be capable of supporting larger keys and more complex cryptographic operations **[41][42]**. As NIST is still finalizing its post-quantum cryptography standards, organizations need to be flexible to adopt future recommendations. Research highlights that this standardization process should influence the design of scalable infrastructure to accommodate future cryptographic updates **[41]**.

*d) Software Libraries Integration*

Research has been focusing on developing secure libraries for implementing NIST's PQC algorithms. Ensuring that widely used libraries like OpenSSL, BoringSSL, and other cryptographic libraries integrate PQC algorithms is essential for secure and efficient deployment **[42]**. A significant body of research focuses on integrating post-quantum algorithms into cloud-based platforms. Cloud providers are actively working to provide quantum-resistant encryption as part of their services, ensuring that data stored in virtualized environments remains secure against quantum threats **[42]**.

*e) Key Management*

Post-quantum cryptographic algorithms typically use much larger key sizes than classical systems. Research stresses the importance of updating Key Management Systems (KMS) to handle these larger keys, ensure secure storage, and perform efficient key exchanges. This is especially crucial in distributed systems where key management is a critical bottleneck **[41]**.

*f) New Protocol*

New protocols that incorporate PQC algorithms into secure communication frameworks are necessary. This includes integrating PQC into common protocols like IPSec and TLS. Organizations need to update their communication systems to be compatible with these emerging standards **[42]**.

*2) Infrastructure for STL-2 Strategic Approach*

Recent advancements in the infrastructure for implementing quantum random number generators (QRNG) have seen significant progress, especially in terms of hardware scalability, integration with existing cryptographic systems, and ensuring true randomness for security-critical applications. These developments indicate that QRNGs are moving beyond niche applications into mainstream cryptographic and security systems, providing a quantum-resistant edge to modern digital infrastructure **[43] [44] [45]**.

*a) Device Diminishment & Integration*

Recent research has focused on reducing the size and complexity of QRNGs to make them more compatible with existing devices. Compact, chip-based QRNGs that use photonic circuits have become a leading approach, making them easier to integrate into mobile devices, IoT systems, and cloud-based infrastructure. This miniaturization allows for widespread deployment in everyday consumer electronics, offering a higher degree of security for user data **[42]**.

*b) Cloud-Based and Edge Implementation*

The cloud and edge computing ecosystems are evolving with QRNG integration to enhance security. The use of cloud-based QRNGs ensures that random numbers generated for encryption are truly random, reducing the risks of cryptographic attacks. Research is focused on implementing QRNGs that can operate at the edge of networks, especially for real-time security needs, in areas like financial services and critical infrastructure.

*c) Enhancing Quantum Entropy Source Efficiency*

Quantum entropy sources, critical for the generation of truly random numbers, have seen improved efficiency with new designs that make use of single-photon detectors and advanced quantum optics. The use of these high-efficiency sources ensures that random number generation happens at higher speeds and with lower error rates, making QRNGs more practical for real-world use cases such as cryptographic applications and large-scale secure communications.

*d) Integration with Classical Cryptographic Systems*

QRNGs are increasingly being integrated with classical cryptographic infrastructures, enhancing the security of existing cryptographic algorithms by introducing truly random keys. This hybrid approach combines the robustness of classical algorithms with the unpredictability of quantum randomness, which is crucial in preventing predictability in encryption schemes.

*e) Quantum-Secured Cloud Computing*

A significant trend is the integration of QRNGs into cloud computing environments to offer quantum-secured encryption services. This allows organizations to leverage the true randomness of QRNGs for key generation in secure cloud communications and storage, especially in financial and healthcare sectors that require robust data protection.

*3) Infrastructure for STL-3 Strategic Approach*

Recent research into the infrastructure requirements for implementing quantum key distribution (QKD) highlights several key considerations, particularly for organizations looking to achieve quantum-resistant security at highest level. These infrastructure developments are vital for organizations aiming to adopt QKD technology as part of their broader quantum-resistant strategy **[46] [47]**.

*a) Scalability and Integration with Existing Networks*

Implementing QKD requires significant adaptation to traditional network infrastructures. Quantum communication protocols, especially over Fiber optics, need specialized equipment such as quantum repeaters **[47]** and low-noise detectors to support long-distance communication. For instance, researchers have demonstrated the use of twin-field QKD over distances of up to 1000 km using Fiber optics **[42]**.

*b) Security Module Integration*

The ISO has introduced standards that address the secure integration of QKD modules into conventional network systems. This is particularly important to ensure that QKD systems are protected from conventional network threats, such as denial-of-service attacks. These standards align QKD with cryptographic module testing and certification, thus promoting secure and standardized deployment **[41]**.

*c) Quantum Repeater Technology*

Quantum repeaters are a critical piece of hardware required for extending the range of QKD beyond current Fiber distance limitations. Recent advancements have shown that quantum repeaters are essential for overcoming the distance limitations in Fiber-based QKD and are expected to become central to the widespread deployment of quantum communication systems **[42] [47]**.

*d) Satellite-Based QKD*

Satellite-based QKD has emerged as a viable solution for extending secure communication to global scales. Researchers have successfully demonstrated satellite-to-ground QKD, which enables secure key exchanges over long distances without the need for Fiber optics **[42]**.

*C. Adoption Strategy*

The transition to post-quantum cryptography requires a proactive and strategic approach, especially in terms of hardware management. Small organizations can leverage cloud-based services, while mid-size and large organizations may need to invest significantly in upgrading data centers, HSMs, and network infrastructure to support the computational demands of PQC. The **Fig 8** reflects the proposed adoption strategy.

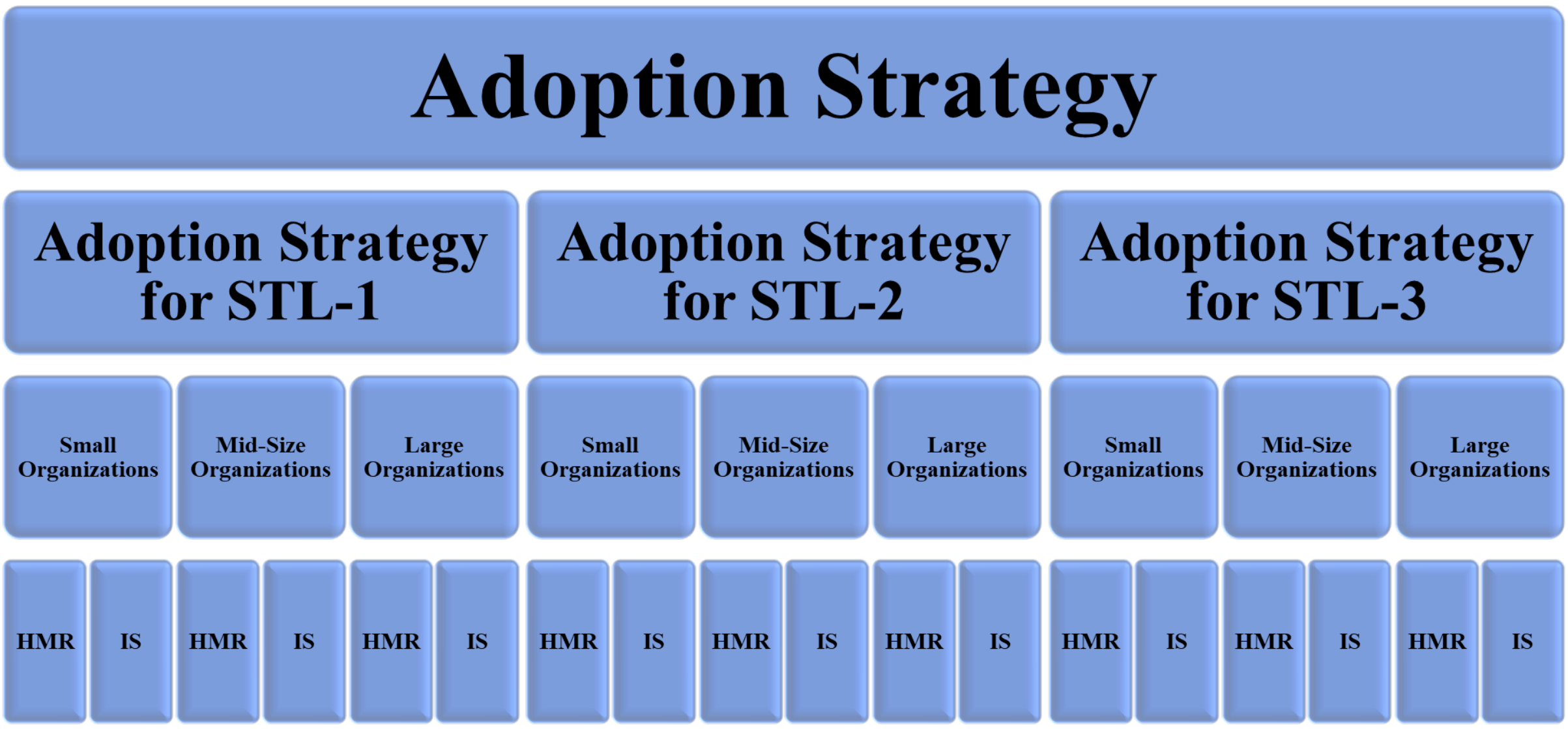

Fig 8: STL-QCRYPTO framework adoption strategy for small, mid-sized and large organizations. Here, STL-X, HMR, and IS represent "Strategic Transition Level X Framework", "Hardware Management Requirements", and "Implementation Strategy" respectively.

*1) Adoption Strategy for STL-1*

*a) Small Organizations*

Hardware Management Requirements

- **Minimal IT infrastructure**: Small organizations typically rely on cloud-based solutions or outsourced IT infrastructure.
- **Low computational resources**: Quantum-resistant algorithms like CRYSTALS-Kyber or FALCON might increase computational overhead on devices like servers or endpoints. Hardware optimization is crucial to handle the additional processing.
- **Basic key management systems**: Hardware security modules (HSMs) or secure enclaves can manage quantum-resistant keys.

Implementation Strategy

- **Cloud-based PQC adoption**: Use cloud providers (like AWS, Azure) that are developing quantum-resistant cryptographic solutions. Small organizations can offload infrastructure costs to these providers.
- **Quantum-safe VPNs and TLS**: Use quantum-safe VPNs for remote access and TLS for web communications.
- **End-to-End Encryption**: Implement PQC algorithms such as CRYSTALS-Kyber for secure communications, using software-based solutions that minimize hardware investment.

For example, A small e-commerce business can start with a PQC upgrade by using a cloud-based HSM service that supports Kyber or Dilithium, ensuring that their data is quantum-resistant without a significant hardware investment.

*b) Mid-Size Organizations*

Hardware Management Requirements

- **Dedicated HSMs and key management**: Mid-size organizations might need to

upgrade or acquire dedicated HSMs to support the more intensive key management demands of PQC.
- **Higher processing capacity**: Quantum-resistant algorithms generally require more computational power. Servers, workstations, and network hardware may need upgrades to ensure sufficient capacity.
- **Legacy system integration**: Transitioning legacy systems to support PQC will require hardware capable of running both classical and quantum-safe algorithms during the migration phase.

Implementation Strategy

- **Hybrid Cryptographic Models**: Implement a hybrid approach that combines classical and post-quantum cryptography to maintain backward compatibility while progressively adopting PQC.
- **Invest in PQC-compatible HSMs**: Ensure that HSMs and encryption modules are capable of supporting post-quantum algorithms such as CRYSTALS-Dilithium and FALCON.
- **Network Layer Security**: Upgrade firewalls, routers, and VPNs to support PQC standards. Mid-size organizations might also need to invest in network hardware that can handle more complex cryptographic operations.
- **Progressive Hardware Upgrades**: Adopt a phased approach to replacing vulnerable hardware with quantum-resistant components, focusing on key systems first.

For example, a mid-size healthcare provider could upgrade its data centers by incorporating PQC-compatible HSMs and network devices that ensure patient records remain secure in the quantum era.

*c) Large Organizations*

Hardware Management Requirements

- **High-performance HSMs and dedicated cryptographic hardware**: Large organizations require robust, scalable HSMs capable of handling quantum-resistant algorithms.
- **Upgrading data center infrastructure**: Data centers must support the increased computational load posed by post-quantum algorithms, such as SPHINCS+ or CRYSTALS-Kyber, which may demand upgrades to processors, storage, and networking equipment.
- **Key management at scale**: Large organizations must implement highly efficient quantum-safe key management systems to handle massive numbers of secure communications and data.

Implementation Strategy

- **Enterprise-Wide PQC Migration Strategy**: Develop a company-wide strategy that involves both hardware and software upgrades, starting with critical systems like financial transactions and sensitive data storage.
- **Quantum-Safe HSMs and Key Management**: Invest in high-capacity HSMs that are purpose-built to support PQC. These HSMs must be compatible with algorithms like CRYSTALS-Kyber for key exchanges and FALCON for digital signatures.
- **Quantum-Classical Hybrid Network Infrastructure**: Build a hybrid infrastructure that combines existing classical cryptographic systems with quantum-safe solutions to ensure backward compatibility and security during the transition period.
- **Data Center Overhaul**: Large-scale hardware upgrades will be necessary to integrate PQC into the data centers, including replacing outdated routers, switches, and network security appliances with quantum-resistant alternatives.
- **Dedicated Quantum Security Teams**: Establish specialized teams focused on managing quantum cryptography systems and coordinating with hardware vendors for post-quantum cryptography readiness.

For example, a multinational financial institution might implement PQC by upgrading its global data centers to support quantum-safe algorithms like Kyber for key exchange and FALCON for secure transactions, ensuring secure international financial transactions.

*2) Adoption Strategy for STL-2*

Implementing STL-2 **Post-Quantum Cryptography (PQC)** with a classical-quantum hybrid approach using **Quantum Random Number Generators (QRNG)** requires different hardware management strategies based on the size and complexity of the organization. The focus is on integrating quantum-safe solutions with classical systems to prepare for quantum computing threats while maintaining efficiency and security.

Hardware management strategies for integrating **STL-2 Post-Quantum Cryptography** with **Quantum Random Number Generators** must be tailored to the scale and complexity of the organization. While small organizations can rely on cloud-based solutions and plug-and-play QRNG devices, mid-size and large organizations will need to invest in dedicated QRNG hardware, scalable HSMs, and hybrid cryptographic systems to ensure post-quantum security. Let's consider some general best practices for all sizes:

- **Vendor Collaboration**: Work with vendors that are developing QRNG-compatible hardware and quantum-resistant cryptography solutions.
- **Security Audits**: Regularly audit existing systems to identify vulnerabilities that could be exploited by quantum attacks, particularly in key management systems.
- **Phased Implementation**: Implement post-quantum cryptography in phases, starting with the most critical systems (e.g., customer databases, financial transactions) and gradually extending to other parts of the organization.

*a) Small Organizations*

Hardware Management Requirements

- **Minimal infrastructure**: Typically rely on cloud-based or outsourced services, so on-premise hardware may be limited.
- **Lightweight QRNG Integration**: QRNG solutions can be implemented through cloud services or low-cost plug-and-play devices that require minimal physical infrastructure.
- **Hybrid Systems**: Classical encryption systems integrated with QRNG for secure key generation and management.

Implementation Strategy

- **Cloud-based QRNG Solutions**: Small organizations should leverage cloud providers that offer quantum-safe encryption as part of their services (e.g., AWS, IBM Quantum).
- **Plug-and-Play QRNG Devices**: For on-premise operations, small businesses can use affordable QRNG devices that integrate into existing systems to enhance key generation without requiring significant hardware investment.
- **Use of Quantum-Safe Protocols**: Implement hybrid cryptographic protocols using classical encryption and quantum-resistant algorithms like CRYSTALS-Kyber.

For example, a small e-commerce platform can upgrade its encryption layer by adopting a cloud-based QRNG service that generates secure, quantum-safe keys for all online transactions without investing in high-end infrastructure.

*b) Mid-Size Organizations*

Hardware Management Requirements

- **Moderate infrastructure**: A mix of on-premise data centers and cloud-based services, with some investment in specialized hardware like servers and encryption devices.
- **QRNG Compatibility with Existing Systems**: Mid-size organizations need to ensure that QRNG devices can be integrated into existing infrastructure without disrupting operations.
- **Key Management**: Implementing advanced key management systems that support classical-quantum hybrid encryption approaches.

Implementation Strategy

- **Dedicated QRNG Hardware**: Invest in QRNG devices to secure sensitive data such as customer financial and healthcare records. Mid-size organizations should also begin incorporating quantum-safe algorithms like CRYSTALS-Dilithium for digital signatures.
- **Hybrid Cloud and On-Premise Systems**: Implement a hybrid infrastructure combining classical cryptography and PQC. QRNG can be integrated into both cloud and on-premise systems to strengthen key generation and cryptographic operations.
- **Data Center Upgrades**: Upgrade existing data centers to handle increased computational loads required for quantum-safe algorithms.

For example, a healthcare provider could deploy dedicated QRNG hardware to secure sensitive patient data while upgrading its key management systems to support a hybrid cryptographic approach, ensuring that both classical and quantum-safe encryption are in place.

*c) Large Organizations*

Hardware Management Requirements

- **Complex, Global Infrastructure**: Large organizations often have extensive on-premise data centers and complex global operations, necessitating more advanced hardware solutions.

- **High-Performance QRNG Systems**: QRNG systems must scale to meet the needs of large organizations, including high-volume transactions, real-time data encryption, and secure communications across global networks.
- **HSM Integration**: Quantum-safe Hardware Security Modules (HSMs) with built-in QRNG capabilities will be essential for managing and distributing cryptographic keys securely across large infrastructures.

<u>Implementation Strategy</u>

- **Enterprise-Grade QRNG Integration**: Implement enterprise-grade QRNG solutions integrated with HSMs and key management systems to ensure secure key generation and distribution across multiple locations and departments.
- **Quantum-Classical Hybrid Encryption**: Adopt a phased approach, starting with quantum-safe encryption for critical systems (e.g., financial transactions, government communications). Use CRYSTALS-Kyber for key exchanges and CRYSTALS-Dilithium for digital signatures, combined with classical encryption in a hybrid model.
- **Scalable Infrastructure**: Upgrade network and data center hardware to handle the additional computational load of PQC algorithms. Ensure that hardware is capable of supporting both classical and quantum operations as part of a long-term strategy for post-quantum security.

For example, a multinational financial institution could implement a global quantum-safe infrastructure by integrating QRNG-powered HSMs across all data centers, ensuring secure key management for large-scale, cross-border transactions.

*3) Adoption Strategy for STL-3*

**Level 3 Post-Quantum Cryptography** strategy integrates **Quantum Key Distribution (QKD)** with **Quantum Random Number Generators (QRNG)** to enhance at the highest level of secure encryption capabilities. This strategic level focuses on providing maximum security for organizations against post-quantum attacks by using the principles of quantum mechanics to generate and distribute cryptographic keys. Hardware management strategies for implementation vary depending on organizational size.

The implementation of **STL-3 post-quantum cryptography** with **QKD** and **QRNG** requires tailored hardware management strategies depending on organizational size. Small organizations can leverage cloud-based QKD and QRNG solutions, mid-size organizations may need to invest in dedicated QKD hardware, and large organizations must build a global QKD network with enterprise-grade QRNG systems. Each strategy should ensure the security of key generation, storage, and distribution to protect sensitive data and communications from future quantum threats. The general best practices for all sizes can consider the following:

- **Vendor Collaboration**: Collaborate with vendors providing quantum-safe infrastructure for QKD and QRNG services.
- **Phased Implementation**: Implement QKD and QRNG in stages, starting with critical data and communications systems, and gradually expanding across the organization's IT infrastructure.
- **Security Audits and Upgrades**: Regularly audit existing infrastructure to ensure compatibility with QKD hardware and QRNG devices and identify areas requiring upgrades or additional security measures.

*a) Small Organizations*

<u>Hardware Management Requirements</u>

- **Minimal Infrastructure**: Small organizations often lack complex in-house IT infrastructure, relying on third-party or cloud-based solutions.
- **Cloud-Based QKD Services**: Instead of investing in expensive QKD hardware, small organizations should leverage cloud-based QKD services provided by vendors.
- **Plug-and-Play QRNG Devices**: Basic, low-cost QRNG hardware devices that can easily integrate with existing systems.

<u>Implementation Strategy</u>

- **Cloud QKD Integration**: Small organizations can utilize managed quantum services from vendors that offer QKD as a service. This allows encryption keys to be securely generated and distributed without requiring costly in-house hardware.
- **QRNG Solutions for Enhanced Security**: Use QRNG-based hardware tokens or small-scale QRNG solutions to enhance the randomness and security of key generation in critical systems (e.g., e-commerce, secure communication channels).

- **Focus on Critical Systems**: Implement QKD for the most sensitive data and communications, while relying on classical post-quantum algorithms for other parts of the system.

For example, a small online business could leverage QKD cloud services to secure payment transactions without significant capital expenditure, while QRNG solutions ensure truly random keys for encrypted communication.

*b) Mid-Size Organizations*

Hardware Management Requirements

- **Moderate Infrastructure**: Mid-size organizations often have hybrid IT environments, combining cloud services with on-premise data centers.
- **Dedicated QKD Nodes**: For enhanced security, organizations might need to invest in QKD hardware devices that integrate into their existing network.
- **QRNG Integration for Enhanced Encryption**: More robust QRNG devices are needed to secure critical data such as customer information, healthcare data, or financial transactions.

Implementation Strategy

- **Deploy QKD in Key Areas**: Implement QKD in systems handling highly sensitive information, such as customer databases, financial transactions, or healthcare records. These systems should be integrated with QRNG-based encryption solutions.
- **Data Center Upgrades**: Organizations should upgrade data center hardware to support QKD nodes, allowing quantum keys to be securely distributed across their global network.
- **Hybrid Cryptography Models**: Combine classical post-quantum cryptographic algorithms with QKD for a hybrid security approach, using QRNG for key generation in both classical and quantum encryption layers.

For example, a mid-size healthcare provider could implement QKD in their data centers for patient record security, while deploying QRNG-enhanced encryption for all communications and cloud storage.

*c) Large Organizations*

Hardware Management Requirements

- **Complex IT Infrastructure**: Large organizations manage extensive data centers, global operations, and multi-location IT systems. This requires advanced hardware solutions for implementing QKD.
- **Enterprise-Level QKD Infrastructure**: Large-scale QKD solutions that secure global communications across multiple data centers and locations.
- **QRNG Hardware Integrated with HSMs**: Quantum-safe Hardware Security Modules (HSMs) integrated with QRNG for secure key management and distribution across complex infrastructures.

Implementation Strategy

- **Global QKD Network**: Large organizations should implement a global QKD network to ensure secure communication across regions and departments. This would require deploying QKD hardware at major data centers and branch offices.
- **Enterprise-Grade QRNG Systems**: Implement enterprise-grade QRNG systems integrated with HSMs for secure key management and storage, ensuring that cryptographic keys are quantum-safe across all operations.
- **Compliance and Regulatory Focus**: For large organizations in highly regulated industries, such as financial services or defence, QKD and QRNG can meet compliance and security mandates, ensuring data is protected from future quantum threats.

For example, a global financial institution with offices worldwide can use a QKD infrastructure to secure inter-office communications and transactions, while QRNG-based HSMs manage cryptographic keys in their global data centers.

# V. Impact Assessment: Industry Sectors at Risk

The imminent threat posed by post-quantum attacks will disrupt industries across the board, necessitating a comprehensive approach to risk mitigation. To effectively safeguard against these emerging vulnerabilities, businesses must first conduct thorough assessments to identify critical gaps in their current security frameworks. These assessments will inform the development of a quantum attack mitigation strategy that aligns with both immediate and long-term needs. A phased, strategic implementation plan across high-impact industries, including financial services, healthcare, and government etc. is essential. This approach involves strategic STL-1 solutions using post-quantum cryptographic algorithms like CRYSTALS-Kyber, which can be integrated into existing classical systems. At STL-2, a hybrid quantum-classical strategy incorporating Quantum Random Number Generators (QRNGs) will enhance unpredictability in encryption, creating stronger defence layers. Finally, STL-3 employs Quantum Key Distribution (QKD), providing an unbreakable quantum-safe method of key exchange to protect highly sensitive data and communications. By advancing through these stages, organizations can ensure that their systems remain secure in an era of quantum computing advancements while future-proofing their infrastructures. We are discussing the following key industries (**Fig 9**):

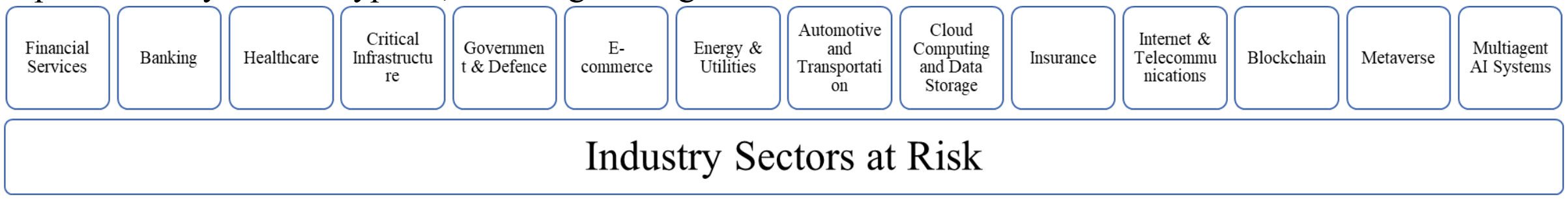

Fig 9: Fourteen key industries identified as being at significant high risk in a post-quantum world

### *A. Financial Services*

#### *1) Aligned with STL-1*

<u>Critical Application Area:</u>

International fund transfers and high-value transactions are particularly vulnerable to advanced cyber threats, including those posed by quantum computers.

<u>Viable solution methodology:</u>

- CRYSTALS-Kyber: Use Kyber for securing cross-border payment systems to ensure that transaction data is encrypted using quantum-resistant algorithms. Kyber is efficient in both encryption and decryption, making it suitable for large-scale systems like SWIFT or other international financial networks.
- CRYSTALS-Dilithium: For digital signatures in international fund transfers, Dilithium can ensure that transaction validation and authentication remain quantum-proof. This would provide long-term security for the financial records associated with cross-border payments.

#### *2) Aligned with STL-2*

<u>Critical Application Area:</u>

Large-scale financial transactions and automated trading platforms depend on encryption protocols to protect client data and trading strategies from unauthorized access.

<u>Viable solution methodology:</u>

- QRNGs for Encryption: Financial services can implement QRNGs to generate truly random encryption keys for financial transactions, providing quantum-resilient security. By integrating QRNGs into current encryption methods (e.g., Advanced Encryption Standard), the financial industry can ensure unpredictability and avoid predictable patterns that adversaries could exploit using classical or quantum attacks.
- Quantum Key Distribution (QKD): Financial institutions can employ QKD using QRNGs to establish secure, unbreakable communication channels **[48]**, ensuring that encryption keys cannot be intercepted or cracked by quantum computers.

#### *3) Aligned with STL-3*

<u>Critical Application Area:</u>

High-frequency trading, cross-border payments, and data-sensitive financial communications are vulnerable to cyberattacks.

<u>Viable solution methodology</u>: QKD for Transaction Security: Quantum Key Distribution can be used to encrypt high-value transactions in real-time, ensuring that any attempt to intercept the encryption keys will be detected immediately. This guarantees the integrity of cross-border payments and secures communication channels between financial institutions. JPMorgan Chase **[49]** has experimented with QKD to safeguard financial transaction communications from quantum threats.

### *B. Banking*

#### *1) Aligned with STL-1*

<u>Critical Application Area:</u>

Online banking services rely heavily on secure authentication protocols to safeguard user identities and transactions.

<u>Viable solution methodology:</u>

- SPHINCS+: This hash-based signature scheme can be applied to multi-factor authentication systems in online banking, ensuring that digital signatures cannot be tampered with, even by a quantum computer. SPHINCS+ is ideal for use in systems that require high security and long-term integrity for sensitive banking operations.
- FALCON: FALCON's small signature size and speed make it an ideal fit for real-time banking authentication, especially for mobile banking applications where efficiency and security must be balanced. FALCON could secure the validation of login sessions and ensure that even if intercepted, they cannot be decrypted by quantum threats.

*2) Aligned with STL-2*

Critical Application Area:

Authentication in online and mobile banking requires robust, unpredictable mechanisms to verify user identities and ensure transaction safety.

Viable solution methodology:

- QRNGs for OTP Generation: Banks can use QRNGs to generate One-Time Passwords (OTPs) for two-factor authentication (2FA) systems. True randomness from QRNGs ensures that OTPs are entirely unpredictable and immune to quantum or classical attacks, securing customer accounts against fraud **[50]**.
- Quantum-Enhanced Public Key Infrastructure (PKI): Banks can employ QRNGs to secure their PKI, ensuring that keys used for digital signatures and certificate issuance are resistant to quantum decryption.

*3) Aligned with STL-3*

Critical Application Area:

Interbank communication, customer data exchanges, and secure vault operations.

Viable solution methodology:

Banks can implement QKD to secure communications between branch offices, central servers, and ATMs. This ensures that sensitive information, such as customer data and transaction logs, are protected from eavesdropping by quantum-capable adversaries. Example: Banks in Switzerland have explored using QKD for SWIFT communications between banks **[51]**.

*C. Healthcare*

*1) Aligned with STL-1*

Critical Application Area:

Healthcare organizations are increasingly reliant on digital platforms to store and exchange sensitive medical data, making them prime targets for future quantum-based attacks.

Viable solution methodology:

- CRYSTALS-Kyber: Kyber can be integrated into healthcare cloud platforms to encrypt EHRs and medical communications. Its efficiency allows secure storage and transmission of sensitive data, ensuring patient confidentiality even in the face of quantum attacks.
- CRYSTALS-Dilithium: For signing and verifying medical documents, Dilithium ensures that digital signatures remain secure and cannot be falsified, providing quantum-safe authentication for both patients and healthcare providers.

*2) Aligned with STL-2*

Critical Application Area:

Healthcare providers need to secure patient data, including electronic health records (EHRs), clinical trials, and communications between medical devices and doctors.

Viable solution methodology:

- QRNGs for Medical Data Encryption: QRNGs can generate secure encryption keys to protect sensitive healthcare data. This ensures that patient information cannot be intercepted or decrypted by quantum computers. Healthcare institutions can use QRNG-based encryption protocols to protect EHRs in transit and storage.
- QKD for Medical Communications: Secure communication between medical devices and health systems using QKD ensures that encryption keys used in data transmission cannot be intercepted, thus safeguarding sensitive health data from future quantum threats.

*3) Aligned with STL-3*

Critical Application Area:

Protecting patient data, securing electronic health records (EHRs), and ensuring privacy in telemedicine.

Viable solution methodology:

QKD for Medical Data Encryption: Hospitals can leverage QKD to protect EHRs and other sensitive patient information from being compromised. Quantum-safe key distribution will secure communications between healthcare facilities, ensuring that medical data remains confidential even against quantum attacks. Example: A hospital could implement QKD to secure transmissions between diagnostic machines, such as MRI scanners, and central medical databases, making sure patient data is quantum-safe **[52]**.

*D. Critical Infrastructure*

*1) Aligned with STL-1*

Critical Application Area:

Smart grids and critical infrastructure systems rely on secure communication between distributed assets and control centers to ensure uninterrupted and safe service delivery.

Viable solution methodology:

- CRYSTALS-Kyber: Kyber can be applied to encrypt communication channels within smart grid systems, ensuring that all data transmitted between sensors, meters, and control systems is resistant to quantum attacks. Its ability to operate efficiently in low-latency environments is ideal for real-time monitoring of critical infrastructure.
- FALCON: In the case of verifying grid commands and updates from control centers, FALCON's small signature size and fast verification times make it ideal for maintaining the security of distributed systems while keeping latency low.

*2) Aligned with STL-2*

Critical Application Area:

Industrial control systems (ICS) in critical infrastructure, such as energy grids, water systems, and transportation networks, are increasingly vulnerable to cyberattacks.

Viable solution methodology:

- QRNG for Control Command Encryption: QRNGs can be used to generate secure, truly random encryption keys for communication between sensors and control systems in smart grids. By integrating QRNGs with traditional ICS, infrastructure operators can prevent cyberattacks and ensure that adversaries cannot predict or decrypt control commands.
- QKD for Control Centers: By using QKD, critical infrastructure can secure communication between control centers and remote monitoring devices, ensuring that key exchanges are quantum-safe and resistant to interception.

*3) Aligned with STL-3*

Critical Application Area:

Supervisory Control and Data Acquisition (SCADA) systems managing power grids, water systems, and industrial control systems.

Viable solution methodology:

QKD for Infrastructure Protection: QKD can be deployed to secure communication channels between SCADA system components, ensuring that operational data and control commands are exchanged securely. This prevents potential cyberattacks on critical infrastructure, such as tampering with power grids or disrupting water systems. Example: European utility companies have experimented with QKD to secure smart grid communications **[52]**.

*E. Government & Defence*

*1) Aligned with STL-1*

Critical Application Area:

Governments and defence systems must ensure the confidentiality and authenticity of sensitive communications that could be intercepted or tampered with by quantum computers in the future.

Viable solution methodology:

- SPHINCS+: This signature scheme is ideal for signing highly classified documents and communications, offering long-term security against quantum attacks. SPHINCS+ can be integrated into defence messaging systems to ensure data integrity.
- CRYSTALS-Dilithium: Dilithium can be used to verify and authenticate government directives, defence communications, and classified information. Its efficient signature verification is ideal for high-security environments that demand robust quantum resistance without compromising operational efficiency.

*2) Aligned with STL-2*

Critical Application Area:

Governments and defence systems need to secure highly sensitive and classified data that could be compromised by quantum computers.

Viable solution methodology:

- QRNG for Military Communication Systems: QRNGs can generate truly random cryptographic keys for securing defence communication systems, such as satellite communications and tactical radios, ensuring that military communications are immune to quantum cryptanalysis **[53]**.
- QKD for Diplomatic Channels: QKD, powered by QRNGs, can be deployed to protect diplomatic and military communication channels, ensuring that encryption keys exchanged for sensitive information remain secure from interception by hostile quantum-enabled adversaries.

*3) Aligned with STL-3*

Critical Application Area:

National security communications, diplomatic transmissions, and secure military operations.

Viable solution methodology:

QKD for Classified Communication Networks: Governments can use QKD to secure diplomatic channels and defence communication systems, ensuring that any attempts to intercept or decrypt sensitive information are immediately detectable. This would be particularly critical for securing satellite communications, military command-and-control systems, and intelligence transmissions. Example: The Chinese government has already implemented QKD in its Micius satellite project to enable quantum-safe communication between ground stations and satellites **[54][55]**.

*F. E-Commerce*

*1) Aligned with STL-1*

Critical Application Area:

Secure customer payment systems and sensitive data from quantum threats in online transactions.

Viable Solution Methodology:

- CRYSTALS-Kyber: Use Kyber for securing payment transactions and customer data in e-commerce platforms, providing quantum-resistant encryption for credit card details, personal data, and transaction records.
- CRYSTALS-Dilithium: Use Dilithium for digitally signing e-commerce transactions to ensure the integrity and authenticity of online purchases.

*2) Aligned with STL-2*

Critical Application Area:

Enhance encryption unpredictability and make it resilient to attacks using quantum computing.

Viable Solution Methodology:

QRNG for Key Generation: Use QRNG to generate truly random encryption keys for secure session management and data protection in online transactions. This ensures that encryption keys for e-commerce transactions are not predictable or vulnerable to quantum attacks.

*3) Aligned with STL-3*

Critical Application Area:

Ensures the integrity and confidentiality of encryption keys, protecting transaction data from eavesdropping during transmission.

Viable Solution Methodology:

QKD for Backend Systems: Use QKD for secure key distribution between e-commerce servers and financial institutions, particularly for high-value transactions or sensitive customer communications.

*G. Energy & Utilities*

*1) Aligned with STL-1*

Critical Application Area:

Protects critical infrastructure from quantum-enabled cyberattacks targeting grid communication systems.

Viable Solution Methodology:

CRYSTALS-Kyber: Implement Kyber to encrypt smart grid data, securing the transmission of energy usage information between consumers, utility providers, and grid controllers.

*2) Aligned with STL-2*

Critical Application Area:

Ensures robust protection of grid control data and real-time energy distribution information.

Viable Solution Methodology:

QRNG for Smart Grid Encryption: Use QRNGs to generate encryption keys for securing communications between smart meters and central control units. This ensures that the control systems cannot be compromised by attackers predicting key patterns.

*3) Aligned with STL-3*

Critical Application Area:

Safeguards critical energy infrastructure from quantum-enabled espionage or sabotage.

Viable Solution Methodology:

QKD for Grid-to-Grid Communication: Use QKD to secure communication channels between regional power grids to ensure that operational data, such as load balancing and resource allocation, is protected from eavesdropping and manipulation.

*H. Automotive and Transportation*

*1) Aligned with STL-1*

Critical Application Area:

Ensures that quantum-enabled adversaries cannot intercept and manipulate vehicle communications, enhancing safety and trust in autonomous driving systems.

Viable Solution Methodology:

CRYSTALS-Kyber: Secure vehicle-to-infrastructure (V2I) and vehicle-to-vehicle (V2V) communications by encrypting control commands and data transmission between autonomous vehicles and traffic management systems.

*2) Aligned with STL-2*

Critical Application Area:

Prevents quantum-level attacks on vehicle control systems, ensuring data integrity and safety in autonomous vehicle networks.

Viable Solution Methodology:

QRNG for Vehicle Encryption: Use QRNG to generate encryption keys for real-time V2V communications, ensuring that control commands, such as braking or navigation, are securely transmitted and cannot be predicted or exploited.

*3) Aligned with STL-3*

Critical Application Area:

Secures the integrity of real-time vehicle routing, preventing malicious disruptions in autonomous transportation networks.

Viable Solution Methodology:

QKD for Central Traffic Control Systems: Use QKD to protect communication between centralized traffic control systems and autonomous fleets, ensuring that traffic signals, routing instructions, and control data are immune to interception.

*I. Cloud Computing and Data Storage*

*1) Aligned with STL-1*

Critical Application Area:

Protects sensitive cloud data from future quantum threats, ensuring long-term data security.

Viable Solution Methodology:

CRYSTALS-Kyber: Use Kyber to encrypt sensitive data stored in the cloud, ensuring that customer data and business-critical information are secure from quantum attacks.

*2) Aligned with STL-2*

Critical Application Area:

Prevents the compromise of encryption keys, ensuring data integrity and confidentiality in cloud environments.

Viable Solution Methodology:

QRNG for Encryption Keys: Integrate QRNG for generating encryption keys used in securing cloud-based data transmission and storage services, ensuring unpredictable and robust protection of sensitive files.

*3) Aligned with STL-3*

Critical Application Area:

Provides unbreakable security for data transmission in the cloud, securing communications from quantum eavesdropping.

Viable Solution Methodology:

QKD for Secure Data Transmission: Use QKD to protect the transfer of highly sensitive data between cloud data centers and user devices, ensuring encryption keys cannot be intercepted or compromised during transmission.

*J. Insurance*

*1) Aligned with STL-1*

Critical Application Area:

Protects insurance records and personal customer data from future quantum-based cyberattacks.

Viable Solution Methodology:

CRYSTALS-Kyber: Encrypt sensitive customer information, such as policyholder details, claims data, and financial transactions, using Kyber to prevent quantum-enabled breaches.

*2) Aligned with STL-2*

Critical Application Area:

Enhances the unpredictability of encryption, protecting sensitive data from potential threats.

Viable Solution Methodology:

QRNG for Policyholder Data: Use QRNG to generate encryption keys for securing customer data, ensuring that sensitive insurance information remains private and resistant to quantum-enabled attacks.

*3) Aligned with STL-3*

Critical Application Area:

Prevents unauthorized access to claims data and ensures secure communication between all parties involved in insurance processes.

Viable Solution Methodology:

QKD for Secure Claims Processing: Use QKD to secure communication channels between insurance companies, third-party providers, and policyholders during the claims process, ensuring that sensitive claims data and communications remain secure.

*K. Internet & Telecommunications*

*1) Aligned with STL-1*

Critical Application Area:

Internet and telecommunications providers face increasing risks to secure data transmission as quantum computing evolves, especially in VoIP, VPN, and encrypted communication channels.

Viable solution methodology:

- FALCON: FALCON can be implemented in telecommunications systems to secure digital certificates and session initiations, ensuring quantum-safe authentication for millions of daily interactions. Its efficiency in terms of key generation and signature verification makes it ideal for high-volume telecommunications services.
- SPHINCS+: For securing long-term digital communications, SPHINCS+ can be used to provide secure, hash-based digital signatures that resist quantum decryption attempts. This ensures long-term protection for telecommunication providers' infrastructure.

*2) Aligned with STL-2*

Critical Application Area:

Telecommunications satellites play a vital role in global communications, transmitting data across large distances.

Viable solution methodology:

QRNG for Satellite Encryption: Satellites can employ QRNGs to generate encryption keys for secure communication links with ground stations and other satellites. Since QRNG-generated keys are unpredictable, they offer a higher level of security compared to traditional methods. This ensures the security of satellite data transmissions, protecting sensitive military, governmental, or corporate communications from unauthorized interception.

*3) Aligned with* STL-3

Critical Application Area:

Securing communication over 5G networks, data centers, and undersea cables.

Viable solution methodology:

QKD for Data Center and Network Security: Telecommunication companies can implement QKD for securing data transmission over fiber-optic networks. By protecting communication links between data centers and across undersea cables, QKD prevents any unauthorized interception of sensitive data. Example: BT has explored QKD as a potential solution for securing 5G networks and undersea communications **[56]**.

*L. Blockchain Applications*

*1) Aligned with STL-1*

Critical Application Area:

Blockchain networks are reliant on cryptographic algorithms to secure transactions and smart contracts. Quantum computers could potentially break existing cryptographic protections.

Viable solution methodology:

- CRYSTALS-Kyber: Kyber can be used to encrypt data in blockchain transactions, providing quantum-safe security for the data that is stored on-chain. Its ability to efficiently handle large-scale encryption makes it ideal for blockchain applications.
- SPHINCS+: SPHINCS+ can be integrated into blockchain systems to ensure that smart contract signatures remain secure, protecting them from quantum-based forgery attacks.

*2) Aligned with STL-2*

Critical Application Area:

Blockchain platforms rely heavily on secure cryptographic protocols to ensure transaction integrity and prevent tampering.

Viable solution methodology:

QRNG for Blockchain Consensus Mechanisms: QRNGs can be used to generate truly random numbers for consensus algorithms in blockchain networks, ensuring that mining and transaction validation processes are unpredictable and immune to quantum attacks.

*3) Aligned with STL-3*

Critical Application Area:

Securing blockchain nodes, validating transactions, and protecting smart contracts.

Viable solution methodology:

QKD for Blockchain Key Distribution: By using QKD, blockchain platforms can secure the cryptographic keys required for validating transactions and running smart contracts. This would ensure that future blockchain applications are resistant to quantum attacks and continue to operate with integrity. Example: There are research initiatives exploring the use of QKD to secure blockchain systems, making blockchain-resistant to future quantum threats **[56]**.

*M. Metaverse Applications*

*1) Aligned with STL-1*

Critical Application Area:

Metaverse platforms rely heavily on the security of digital identities and assets. As quantum computing advances, these virtual identities could be at risk of compromise.

Viable solution methodology:

- CRYSTALS-Dilithium: Dilithium can be applied to verify digital identities in the metaverse, ensuring that user avatars and assets are authenticated securely and cannot be compromised.
- SPHINCS+: This algorithm can ensure that digital assets, including NFTs, are protected from quantum attacks, ensuring their authenticity and ownership remain verifiable in the metaverse.

*2) Aligned with STL-2*

Critical Application Area:

Metaverse platforms need to secure user identities, virtual assets, and transactions within their digital economies.

Viable solution methodology:

QRNG for Metaverse Identity Security: QRNGs can generate secure, random keys to protect digital identities in the metaverse, ensuring that avatars, virtual assets, and user data are secured from quantum attacks.

*3) Aligned with STL-3*

Critical Application Area:

Ensuring the privacy of user data, securing virtual transactions, and protecting digital assets.

Viable solution methodology:

QKD for Identity Protection: Metaverse platforms **[57]** can use QKD to secure user identities and prevent unauthorized access to virtual assets. By using QKD, they can ensure that transactions within the metaverse economy remain protected from quantum threats. Example: Large tech firms are researching the role of QKD in securing digital environments like the Metaverse.

*N. Large and Complex LLM-Based Multiagent AI Systems*

*1) Aligned with STL-1*

Critical Application Area:

In large and complex multiagent systems **[58]** where large language models (LLMs) based agents **[59]** **[60]** are employed, securing communication between agents is critical, especially when sensitive data is involved.

Viable solution methodology:

- FALCON: FALCON can be utilized to secure communication channels between agents in a LLM-based multiagent AI system, providing fast and efficient digital signatures that ensure the integrity and authenticity of inter-agent communications.
- CRYSTALS-Kyber: Kyber can be used for encrypting the data exchanged between agents, ensuring that even if communications are intercepted, they cannot be decrypted by future quantum systems.

*2) Aligned with STL-2*

Critical Application Area:

In systems where large language models (LLMs) based agents are used to coordinate within a multiple agents' eco-system, secure communication is critical.

Viable solution methodology:

QRNG for Agent Communication: Multiagent systems can use QRNGs to secure the communication channels between agents, ensuring that the randomness of communication keys is immune to quantum-based eavesdropping or attacks.

*3) Aligned with STL-3*

Critical Application Area:

Multiagent systems that rely on Large Language Models (LLMs) for decision-making in complex environments.

Viable solution methodology:

QKD for Secure Inter-Agent Communication: Multiagent systems powered by LLMs can use QKD to ensure that communication between agents remains secure and quantum-resistant. This is particularly useful in systems where the agents need to cooperate in real-time, such as autonomous vehicle fleets or smart city management. Example: AI-driven supply chain management systems can use QKD to secure data exchanged between different machine agents, ensuring operational efficiency without data breaches.

# VI. Policy and Regulatory Considerations for Quantum-Resistant Security

The policy and regulatory landscape for quantum-resistant security is in its early stages and is continually evolving. As quantum hardware, software, and related ecosystems develop, policies will need to adapt in response to new advancements. Governments, regulatory bodies, and industry experts are beginning to assess potential quantum vulnerabilities, but current frameworks are limited. To ensure robust security measures, there will be a need for ongoing evaluation and updates, driven by real-world quantum advancements and threat landscapes.

For instance, global organizations like NIST are spearheading efforts to standardize quantum-resistant cryptography, while sectors such as finance, healthcare, and defence will require tailored guidelines to mitigate specific quantum risks. This evolving regulatory framework will also need to consider cross-border collaborations and compliance to ensure a global approach to cybersecurity. Here are key points outlined in the policy and regulatory considerations for quantum-resistant security, reflecting the latest developments and forward-looking initiatives in the field:

## *A. Standardization and Certification*

- **NIST Post-Quantum Cryptography Standardization Project**: NIST is spearheading the standardization of post-quantum cryptographic algorithms. Their current effort focuses on finalizing four primary algorithms like CRYSTALS-Kyber for public-key encryption and CRYSTALS-Dilithium for digital signatures, with anticipated deployment across critical infrastructure and governmental use **[61]**. They have finalized the three standards recently.
- **Certification Programs**: Certification frameworks such as **FIPS 140-3**, which focuses on cryptographic modules, are expected to be updated to integrate quantum-resistant algorithms. These are being discussed within NIST and ENISA frameworks **[62]**.

## *B. Data Protection Regulations*

- **GDPR Compliance with Quantum Threats**: Under the GDPR, organizations will be required to adopt quantum-safe encryption to secure personal and sensitive data against future quantum attacks. This could impact sectors like finance, healthcare, and government **[63]**.
- **CISA's Role**: In the U.S., CISA has been working on guidance for quantum-safe cryptography, focusing on critical sectors such as energy, finance, and defense, which may have to adhere to new quantum-safe standards **[64]**.

## *C. Transition and Interoperability Guidelines*

- **NIST's Hybrid Approach**: NIST recommends adopting a hybrid cryptographic model, combining traditional encryption algorithms with quantum-safe alternatives. This ensures data security during the transition phase while quantum-resistant solutions are fully implemented **[64]**.
- **Interoperability Between Systems**: ENISA and NIST emphasize the importance of ensuring that classical cryptographic systems can work alongside quantum-resistant ones, especially during the multi-year migration **[61]**.

*D. Supply Chain and Critical Infrastructure Security*

- **Quantum-Safe in Critical Infrastructure**: ENISA has outlined that the critical infrastructure sectors—energy, telecommunications, healthcare, and defense—will need to prioritize quantum-safe solutions to protect national security **[61] [64]**.
- **U.S. National Security Agency (NSA) Guidelines**: NSA's cybersecurity efforts have included an active focus on integrating quantum-resistant cryptography in defense and public sector supply chains **[65]**.

*E. Government Mandates and Compliance Deadlines*

- **The U.S. Quantum Computing Cybersecurity Preparedness Act**: This U.S. act mandates that federal agencies assess the vulnerability of current cryptographic systems to quantum attacks and plan a migration path to quantum-safe solutions **[65][63]**.
- **Global Coordination**: International bodies like the G7 and OECD are coordinating efforts to create a harmonized framework for quantum-resilient security across borders **[65][64]**.

*F. Research and Innovation Incentives*

- **Public-Private Partnerships and Funding**: Governments and institutions, particularly in the U.S. and EU, are expected to provide funding and resources for quantum-safe cryptographic research. For example, NIST and ENISA encourage public-private partnerships to expedite the transition **[64]**.
- **Grants for Quantum-Safe Transition**: Governments may provide financial grants to support the adoption of quantum-resistant cryptography, particularly among small and medium-sized enterprises **[66][61]**.

*G. Sector-Specific Regulations*

- **Banking and Finance**: SWIFT, the interbank messaging system, is preparing for quantum-safe cryptography by working with regulatory bodies such as NIST to develop frameworks specific to the financial sector **[64]**.
- **Healthcare**: Regulations such as HIPAA in the U.S. are likely to integrate quantum-safe requirements to ensure the protection of healthcare data **[63]**.

# VII. Implementation Challenges

Both technical and business challenges related to quantum cryptography implementation require strategic planning, incremental adoption, and investment in infrastructure and human capital.

*A. Technical Challenges*

We may face many difficulties encountered during implementing secure quantum communication systems. These include integrating quantum hardware with classical infrastructure, and addressing scalability and cost concerns. Additionally, developing standardized protocols and error correction techniques are key hurdles in ensuring reliable and practical use of quantum cryptography.

*1) Technical Challenges for Implementing STL-1*

*a) Increased Computational Load*

- **Challenge**: PQC algorithms, such as those recommended by NIST (e.g., CRYSTALS-Kyber, Dilithium), typically require more computational resources compared to classical cryptographic systems. This includes increased processing time for key generation, encryption, and decryption, which can significantly slow down systems, especially in real-time applications.
- **Mitigation Plan**: Organizations should gradually upgrade their computational infrastructure to handle the higher demands. A hybrid approach, integrating both classical and post-quantum cryptographic methods during the transition period, will allow systems to adapt without sacrificing performance. Optimizing current hardware (such as integrating more efficient processors) and adopting cloud-based quantum-safe solutions can also help balance performance.

*b) Key Size and Data Overhead*

- **Challenge**: PQC algorithms, such as FALCON and CRYSTALS-Dilithium, typically generate much larger cryptographic keys and ciphertexts than their classical counterparts. This can result in increased data overhead, impacting bandwidth, storage, and communication systems, especially in applications that rely on high data throughput.
- **Mitigation Plan**: Mitigating this issue involves improving storage and network capacities. For mid-to-large-sized organizations, investing in scalable cloud infrastructure with dynamic data management capabilities is essential. Advanced data compression techniques and efficient key management systems can also be implemented to reduce the burden of larger key sizes.

c) *Hardware Compatibility*

- **Challenge**: The current hardware (e.g., Hardware Security Modules, routers, and servers) may not be compatible with the new PQC algorithms due to increased key sizes and processing requirements. Legacy systems may require complete overhauls, which is both costly and time-consuming.
- **Mitigation Plan**: Gradual integration of quantum-safe hardware through a phased approach is critical. For instance, HSMs should be upgraded to support PQC algorithms. Organizations should focus on upgrading key components (e.g., processors, cryptographic accelerators) while maintaining a hybrid cryptographic system during the transition.

2) *Technical Challenges for Implementing STL-2*

a) *Integration with Legacy Systems*

- **Challenge**: QRNG technology may not seamlessly integrate with legacy cryptographic systems that were designed for classical random number generation methods.
- **Mitigation Plan**: Conduct phased upgrades of cryptographic libraries, ensuring backward compatibility during initial implementations. Consider hybrid systems where QRNG is initially used for critical processes only.

b) *Performance and Latency Issues*

- **Challenge**: QRNG can introduce latency in real-time systems, especially when incorporated into high-throughput environments that rely on low-latency encryption.
- **Mitigation Plan**: Optimize the data flow by strategically placing QRNG devices at critical points in the infrastructure. Combine with high-performance cryptographic hardware, such as Hardware Security Modules (HSMs), to reduce delays.

c) *Hardware Reliability and Scalability*

- **Challenge**: QRNGs require specialized quantum hardware that may not be as reliable or scalable as classical random number generators. Small- to mid-size organizations might face challenges due to high costs and technical complexity.
- **Mitigation Plan**: Focus on cloud-based QRNG solutions that provide on-demand access to quantum-secure random numbers, reducing upfront hardware investments. Partner with established quantum service providers to ensure scalability.

3) *Technical Challenges for Implementing STL-3*

a) *Limited Distance and Signal Attenuation*

- **Challenge**: QKD relies on transmitting quantum bits (qubits) over optical fibers or free space. The signal weakens over long distances, limiting its effectiveness to about 100-150 km without quantum repeaters.
- **Mitigation Plan**: Develop and integrate quantum repeaters or satellite-based QKD solutions to extend communication range. Collaboration with research institutions working on quantum networks can accelerate this advancement. A report by the European Telecommunications Standards Institute (ETSI) **[67]** highlights distance limitations and the need for quantum repeaters.

b) *Integration with Existing Classical Infrastructure*

- **Challenge**: QKD requires specialized hardware like quantum-enabled transceivers and cannot easily be integrated with classical cryptographic systems. Compatibility issues arise when combining quantum and classical systems.
- **Mitigation Plan**: Implement a hybrid cryptographic framework that combines classical encryption techniques with QKD for critical data exchange. Gradual upgrades to infrastructure and leveraging cloud-based quantum services will ensure smoother integration.

*c) Security Vulnerabilities in Implementation*

- **Challenge**: While QKD promises theoretically unbreakable security, real-world implementations can introduce vulnerabilities, such as side-channel attacks, which exploit imperfections in the system.
- **Mitigation Plan**: Regularly audit and monitor QKD systems for potential side-channel vulnerabilities. Invest in advanced hardware design that mitigates such risks, and partner with security-focused quantum service providers for deployment.

## *B. Business-Oriented Challenges*

Business-oriented challenges in quantum cryptography involve the high costs of adopting quantum technology, the complexity of integrating it into existing infrastructure, and the lack of skilled professionals in the field. Additionally, organizations face uncertainty due to evolving standards, making it difficult to ensure long-term investment security and regulatory compliance.

### *1) Business-Oriented Challenges for Implementing STL-1 Strategy*

*a) Cost of Transition*

- **Challenge**: The cost of upgrading existing infrastructure and retraining staff to implement PQC can be substantial. This is particularly burdensome for small-to-mid-sized organizations with limited IT budgets.
- **Mitigation Plan**: A staged transition plan should be developed, starting with high-priority, sensitive systems. Businesses can seek government grants or industry partnerships for funding assistance and focus on open-source PQC solutions to minimize costs. Cloud providers offering post-quantum solutions on a subscription basis can also reduce the financial burden.

*b) Lack of Skilled Workforce*

- **Challenge**: PQC is a relatively new and evolving field. Organizations may lack internal experts who understand both the technical and business implications of implementing PQC. The gap in knowledge can delay the adoption and increase operational risks.
- **Mitigation Plan**: Businesses should invest in training and professional development programs to upskill their existing IT teams. Partnerships with academic institutions or hiring specialized consulting firms can also bridge the knowledge gap. Ensuring that key personnel are certified in quantum cryptography and related technologies will streamline the adoption process.

*c) Regulatory Uncertainty*

- **Challenge**: While governments and international bodies like NIST are in the process of finalizing PQC standards, regulatory frameworks are still under development. The absence of clear, unified regulations can hinder organizations from adopting PQC, especially in highly regulated sectors like finance and healthcare.
- **Mitigation Plan**: Organizations should proactively monitor the evolving regulatory landscape and participate in industry-standardization forums to stay ahead of new developments. Establishing internal compliance frameworks based on best practices (e.g., NIST's recommendations) will ensure that businesses are prepared when regulations become more formalized.

### *2) Business-Oriented Challenges for Implementing STL-2*

By addressing following challenges strategically, organizations can gradually implement QRNG while minimizing disruptions and costs.

*a) Cost of Implementation*

- **Challenge**: Implementing QRNG technology is expensive, especially for smaller organizations. The cost includes hardware procurement, system integration, and training.
- **Mitigation Plan**: Leverage quantum cloud service providers to reduce direct hardware investment. Consider gradual deployment in stages to spread out costs over time. Cost challenges are frequently addressed in reports from Quantum Computing Inc. and McKinsey on quantum technology deployment **[68]**.

*b) Market and Regulatory Uncertainty*

- **Challenge**: The regulatory environment around quantum technologies, including QRNG, is still evolving. Lack of clear standards for QRNG deployment creates uncertainty for businesses.
- **Mitigation Plan**: Engage with regulatory bodies and industry working groups to stay updated on quantum security guidelines.

Advocate for clearer standards in post-quantum cryptography and QRNG adoption.

*c) Lack of Skilled Talent*

- **Challenge**: The deployment of QRNG requires expertise in both quantum technologies and classical cryptography, which are niche fields with a limited talent pool.
- **Mitigation Plan**: Invest in employee training and collaborate with academic institutions to develop talent pipelines. Use consultants or external quantum service providers in the interim.

*3) Business-Oriented Technical Challenges for Implementing STL-3*

By addressing these challenges with proactive strategies, organizations can begin to explore the integration of QKD technology while minimizing potential risks and costs.

*a) High Cost of Deployment*

- **Challenge**: The equipment and infrastructure required for QKD, including quantum transceivers and optical fiber networks, are expensive, particularly for small and mid-size enterprises.
- **Mitigation Plan**: Adopt a phased deployment strategy, starting with critical business processes. Partner with cloud providers offering QKD as a service to avoid upfront capital expenditure on hardware. Gartner's 2024 Quantum Security report **[69]** outlines the financial hurdles of quantum technology deployment.

*b) Lack of Industry-Wide Standards and Regulatory Clarity*

- **Challenge**: The absence of globally accepted standards for QKD poses a challenge for businesses in adopting the technology, as it may not comply with future regulatory frameworks.
- **Mitigation Plan**: Stay engaged with regulatory bodies such as NIST and ETSI, which are developing QKD standards. Participate in quantum security consortiums to shape industry standards and ensure future compliance.

*c) Talent Shortage and Skill Gaps*

- **Challenge**: Implementing QKD requires specialized expertise in both quantum physics and cryptography. The shortage of qualified professionals can hinder deployment and maintenance.
- **Mitigation Plan**: Invest in internal training programs and collaborate with universities to develop talent pipelines. In the short term, engage with consultants or quantum technology firms to fill skill gaps during early implementations.

# VIII. Conclusion

In conclusion, quantum computing represents both an extraordinary opportunity and a formidable challenge for industries worldwide. As we advance into a future driven by quantum technologies, the potential to revolutionize sectors such as artificial intelligence, materials science, and healthcare is undeniable. However, alongside these breakthroughs lies the critical issue of cybersecurity. The ability of quantum computers to break classical encryption methods like RSA and other cryptographic algorithms poses a direct threat to the integrity of sensitive data and digital infrastructure. The timeline for the full realization of quantum computing's power may still be uncertain, but the urgency of addressing its security implications is not.

Our discussed strategic framework provides a practical roadmap for industries to begin preparing for the quantum threat today. By categorizing mitigation strategies into three levels—Foundational, Intermediate, and Advanced—our approach enables organizations to prioritize based on their readiness, resource availability, and the complexity of the solutions required. These levels offer a flexible yet structured pathway, ensuring that even foundational steps can enhance quantum resilience while more advanced approaches provide robust, future-proof defences.

We also highlighted various industrial use cases, demonstrating how these strategies can be applied in real-world scenarios, from securing communication networks to protecting financial transactions. Furthermore, our framework stresses the importance of aligning these technological defences with regulatory requirements, ensuring that organizations not only protect themselves against quantum attacks but also remain compliant with emerging legal and policy frameworks.

As industries move towards a more quantum-aware future, the need for proactive, quantum-safe encryption measures becomes increasingly essential. Waiting until quantum computers become fully functional at scale could expose critical data to significant risks, even if that data appears secure today. By adopting a forward-looking approach, industries can mitigate the threats posed by quantum attacks, safeguarding their digital assets and ensuring long-term security in an era of rapid technological change. Now is the time for organizations to act, laying the foundation for a secure, quantum-resistant future.